\documentclass[
aps,
prb,
preprint,
superscriptaddress,
floatfix
]{revtex4-1}
\usepackage{amsmath}
\usepackage{graphicx}
\usepackage{multirow}
\begin{document}
\title{Magnetism in amorphous carbon}
\author{Yuki Sakai}
\affiliation{Center for Computational Materials, Institute for Computational Engineering and Sciences, The University of Texas at Austin, Austin, Texas 78712, USA}
\author{James R.~Chelikowsky}
\affiliation{Center for Computational Materials, Institute for Computational Engineering and Sciences, The University of Texas at Austin, Austin, Texas 78712, USA}
\affiliation{Department of Chemical Engineering, The University of Texas at Austin, Austin, Texas 78712, USA}
\affiliation{Department of Physics, The University of Texas at Austin, Austin, Texas 78712, USA}
\author{Marvin L.~Cohen}
\affiliation{Department of Physics, University of California at Berkeley, Berkeley, California 94720, USA}
\affiliation{Materials Sciences Division, Lawrence Berkeley National Laboratory, Berkeley, California 94720, USA}

\date{\today}

\begin{abstract}
We investigate magnetism in amorphous carbon as suggested by the recently reported ferromagnetism in a new form of amorphous carbon. 
We use spin constrained first-principles simulations to obtain amorphous carbon structures with the desired magnetization.
We show that the existence of $sp^2$-like 3-fold coordinated carbon atoms plays an important role in obtaining magnetism in amorphous carbon.
The detailed geometries of 3-fold carbon atoms induce the magnetic order in amorphous carbon.
\end{abstract}

\maketitle

\section{Introduction}
Diamond and graphite, which are abundant allotropes of carbon, are diamagnetic materials owing to their orbital diamagnetism.\cite{Makarova2004,Esquinazi2005}
However, other types of magnetism in carbon materials have been investigated.
For example, nano-scale graphene nanoribbons are predicted to exhibit magnetic order coming from localized edge electronic states \cite{Fujita1996,Nakada1996,Son2006PRL,Son2006Nature} although experimentally no direct observation of this prediction has been reported in the literature.\cite{Magda2014,Ruffieux2016}
In its bulk form, graphite is reported to exhibit ferromagnetism when irradiated with high energy protons,\cite{Esquinazi2003} when a network of point defects due to grain boundary appears,\cite{Cervenka2009} or vacancies are introduced.\cite{Yazyev2007,Yazyev2010,Ugeda2010}
Theoretical studies have shown that carbon nanotubes can also be magnetic when line defects are introduced, \cite{Okada2006,Alexandre2008} when nanotubes form composites with other nanotubes,\cite{Park2003} or when a graphene-nanotube complex are created under pressure.\cite{Batista2014}

Recently, a new amorphous form of carbon (\textit{Q-carbon}) has been reported as a room-temperature ferromagnetic phase of carbon \cite{Narayan2015,Bhaumik2018}.
The reported magnetic moment is 0.4~$\mu_B$/atom (where $\mu_B$ is the Bohr magneton) with a Curie temperature of 500~K.
Q-carbon exhibits superconductivity when it is born doped owing to the large proportion (75-85~\%) of $sp^3$-hybridized carbon atoms \cite{Bhaumik2017,Bhaumik2017-2,Sakai2017}.
Ferromagnetism in amorphous-like carbon nanofoams has been reported,\cite{Rode2004,Arcon2006} but the magnetization and the fraction of $sp^3$-hybridized carbon atoms is significantly larger in Q-carbon.
Therefore, a theoretical understanding of the ferromagnetism in amorphous carbon may allow us to understand the magnetic and structural characteristics in Q-carbon, and finding a promising alternative to rare-earth magnets is technologically important.

We perform a computational investigation of magnetic amorphous carbon.
A fixed magnetization on carbon atoms is imposed as we construct a model structure of amorphous carbon from liquid-like carbon. 
The spin constrained structure tends to have more 3-fold (near $sp^2$) coordinated carbon atoms than those without spin constraints, indicating the importance of unpaired electrons for obtaining magnetic amorphous carbon.
We also study the effect of the mass density and constrained magnetization on structures and the total energies of amorphous carbon.
Magnetization of order 0.1 to 0.2~$\mu_B$/atom does not yield high energy structures when compared with nonmagnetic cases particularly in low density amorphous carbon, although high energy structures are required to have the experimentally measured magnetization (0.4~$\mu_B$/atom)
Finally we release the magnetic constraint and find that some spin magnetic moments are retained.
The possible magnetic order among these remaining spins are discussed.

\section{Computational Method}
We employ a total energy pseudopotential approach with both
Troullier-Martins norm-conserving pseudopotentials and Vanderbilt ultrasoft pseudopotentials \cite{Cohen1982,Ihm1979,Louie1982,Vanderbilt1990,Troullier1991} 
constructed within density functional theory (DFT) \cite{Hohenberg1964,Kohn1965} 
using the Perdew-Burke-Ernzerhof exchange-correlation functional.\cite{Perdew1996} 
The real-space pseudopotential DFT code PARSEC is used for molecular dynamics (MD) simulations.\cite{Chelikowsky1994,Chelikowsky2000,Kronik2006,Natan2008} 
The plane-wave DFT package Quantum ESPRESSO \cite{Giannozzi2009} is used for performing spin-constrained structural relaxation.
A real-space grid of 0.3~Bohr (1~Bohr $\simeq$ 0.52918~{\AA}) and a plane-wave energy cutoff of 65~Ry are used to obtain sufficiently converged total energies.
Only the $\Gamma$ point is sampled for a Brillouin-zone integration.

MD simulations are performed to construct a model structure of amorphous carbon.
First we prepare a 216~carbon atom supercell in a simple cubic structure.
Next we increase the system temperature to 7500~K and perform MD simulation at 7500~K in an NVT ensemble to randomize the atomic coordinates. 
The temperature is controlled by using Langevin thermostat with a friction constant of 10$^{-3}$~a.u.
Finally we stop the simulation at 500~MD step (time step $\Delta t$ of 1~fs) and relax the atomic coordinates of that step accordingly.

The parameters needed for obtaining amorphous carbon structures are the density and magnetization.
The density of amorphous carbon is adjusted by tuning the lattice parameter of the cubic supercell.
As for the magnetization, we fix the total magnetic moment of the system to a certain value while we perform structural relaxation.
These constrained magnetization calculations are performed by imposing two different Fermi energies for spin up and down electrons as implemented in Quantum ESPRESSO.
We choose zero, 0,1, 0.2, and 0.4 $\mu_B$/atom for magnetic constraints.

\section{Results and Discussion}
\subsection{Dependence on magnetization}
\begin{figure}[tbp]
  \includegraphics[width = 8.6cm]{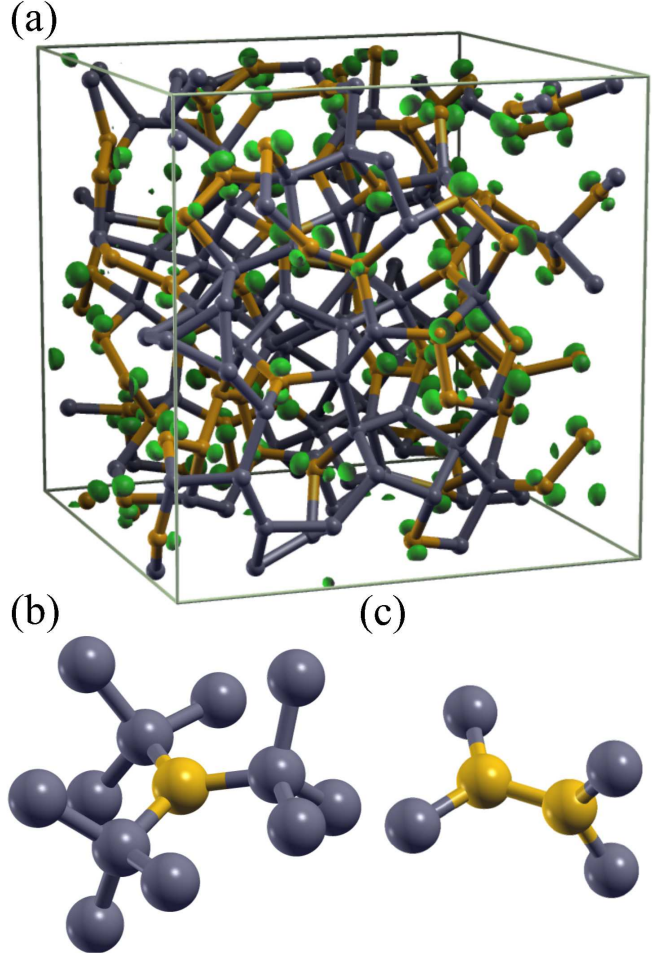}
    \caption{
    \label{fig:fig1}
(Color online) (a) Structure and spin charge density of amorphous carbon with constrained magnetization of 0.4~$\mu_B$/atom and density of 3.4~g/cm$^3$. Gray and orange spheres represent 4-fold and 3-fold coordinated carbon atoms, respectively. Here the bonding threshold distance between two carbon atoms is 1.8~{\AA}. A green isosurface illustrates the spin charge density. Schematic illustrations of (b) a 3-fold carbon atom surrounded by 4-fold atoms and (c) two 3-fold atoms bonded but their $p$ orbitals are rotated relative to each other. 
}
\end{figure}

First we analyze a specific density case of 3.4~g/cm$^3$ (corresponding to a cubic cell with a lattice parameter of 20.445~Bohr) to observe the effect of magnetic constraints on the structure of amorphous carbon. 
Figure~\ref{fig:fig1}(a) shows the relaxed structure and the spin charge density of 0.4~$\mu_B$/atom constrained magnetization case, corresponding to the experimentally-measured magnetic moment.\cite{Narayan2015}
Here 44~\% of carbon atoms are 3-fold coordinated.
These 3-fold carbon atoms (orange spheres) exhibit spin polarization (green isosurface).
Virtually no spin density can be found around 4-fold atoms (illustrated by gray spheres).
Figure~\ref{fig:fig1}(a) qualitatively shows that unpaired electrons associated with 3-fold coordinated atoms are required to induce magnetism in amorphous carbon.

In Fig.~\ref{fig:fig1}(a), one can see two types of 3-fold atoms with unpaired electrons.
The first type of 3-fold atoms are surrounded by three 4-fold coordinated atoms as schematically illustrated in Fig.\ref{fig:fig1}(b).
This creates unpaired electrons since the surrounding 4-fold atoms do not have electrons to form additional bonds with the extra electron at 3-fold atom.
In fact, a model with alternating $sp^2$ and $sp^3$-hybridized carbon atoms was predicted to be ferromagnetic \cite{Ovchinnikov1991}
However, it has been found that such a structure transforms to a more stable phase with less magnetic order when fully relaxed using first-principles methods \cite{Strong2004,Pisani2009}.
Such a separation of $sp^2$ and $sp^3$ hybridized atoms could occur and be a source of magnetic moment in amorphous carbon.

The second type of 3-fold atoms is connected to another 3-fold atom, but still possesses unpaired electrons.
These 3-fold atoms are bonded, but do not form a $\pi$ bond because their extra $p$ electrons are not in parallel [(see Fig.~\ref{fig:fig1}(c)].
Even though we impose a magnetic constraint, neighboring 3-fold carbon atoms forms $\pi$ bond when their unpaired $p$ electrons align in parallel and cannot contribute to the magnetic moment.
Therefore, the relative rotation between two $p$ electrons is necessary for having unpaired electrons, although the energy loss due to this rotation is not negligible as the formation of a $\pi$ bond lowers the energy.
We expect that these structural characteristics could be observed in magnetic amorphous carbon.

\begin{figure}[tbp]
  \includegraphics[width = 8.6cm]{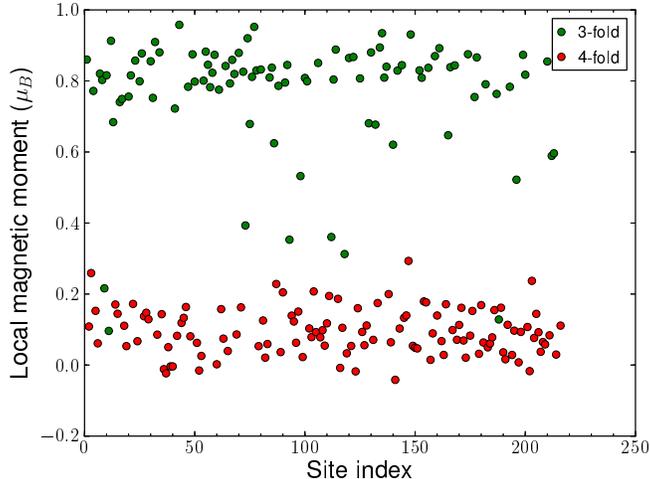}
    \caption{
    \label{fig:pol}
(Color online) Distribution of local magnetic moments at 216 carbon atomic sites in amorphous carbon (0.4~$\mu_B$/atom and 3.4g/cm$^3$). Green circles and red circles represent 3-fold and 4-fold coordinated carbon atoms, respectively. The local magnetic moments are estimated based on the Lowdin charge analysis of spin up and down charge densities. 
}
\end{figure}

Figure~\ref{fig:pol} illustrates the distribution of local magnetic moments on carbon atomic sites with constraint magnetization of 0.4~$\mu_B$/atom (i.e.~86.4~$\mu_B$/cell). 
The 3-fold (green) and 4-fold (red) atomic sites show clear separation in their local moments.
On average, the 3-fold site has approximately 0.8~$\mu_B$/atom while the the moments at 4-fold sites are less than 0.1~$\mu_B$/atom.
Note that some 3-fold sites exhibit small magnetic moments but these moments are artifacts of our use of a supercell and constrained magnetization.
The number of 3-fold sites is 96 and all the 3-fold sites cannot carry magnetic moments to hold the imposed total magnetization of 86.4~$\mu_B$.
Our analysis confirms that unpaired electrons are required to obtain magnetic amorphous carbon. 

\begin{figure}[tbp]
  \includegraphics[width = 8.6cm]{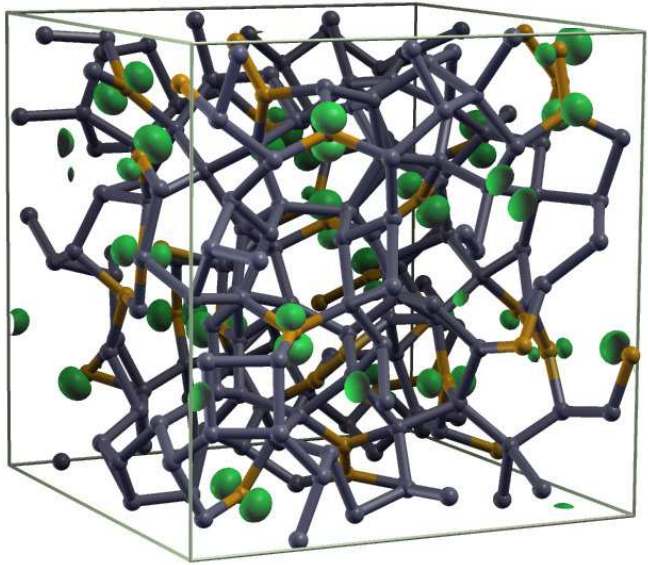}
    \caption{
    \label{fig:diff01}
(Color online) Structure and spin charge density of amorphous carbon with constrained magnetization of 0.1~ $\mu_B$/atom. The density is 3.4~g/cm$^3$.
}
\end{figure}

We also simulate amorphous carbon with a low constrained magnetization to study the change in structure.
By reducing the constraint from 0.4 to 0.1~$\mu_B$/atom, the portion of 3-fold coordinated carbon atoms are reduced from 44~\% to 19~\%.
The reduction of 3-fold carbon atoms can be clearly recognized as the small number of orange spheres in Fig.~\ref{fig:diff01} when compared with those in Fig.~\ref{fig:fig1}.
The spin charge density is still distributed around 3-fold coordinated atoms, but most of these atoms are surrounded by 4-fold carbon atoms owing to the increase (decrease) in 4-fold (3-fold) coordination.
The formation of 3-fold carbon atoms are not favored in amorphous carbon with such a high density of carbon atoms, and it results in the reduction of the 3-fold portion as the magnetization is reduced.
In fact, the total energy is 543~meV/atom higher in the 0.4~$\mu_B$/atom case compared with 0.1~$\mu_B$/atom case, implying the difficulty of the formation of 3-fold atoms in high density amorphous carbon.\cite{Marks1996,Mcculloch2000,Han2007}

\begin{figure}[tbp]
  \includegraphics[width = 8.6cm]{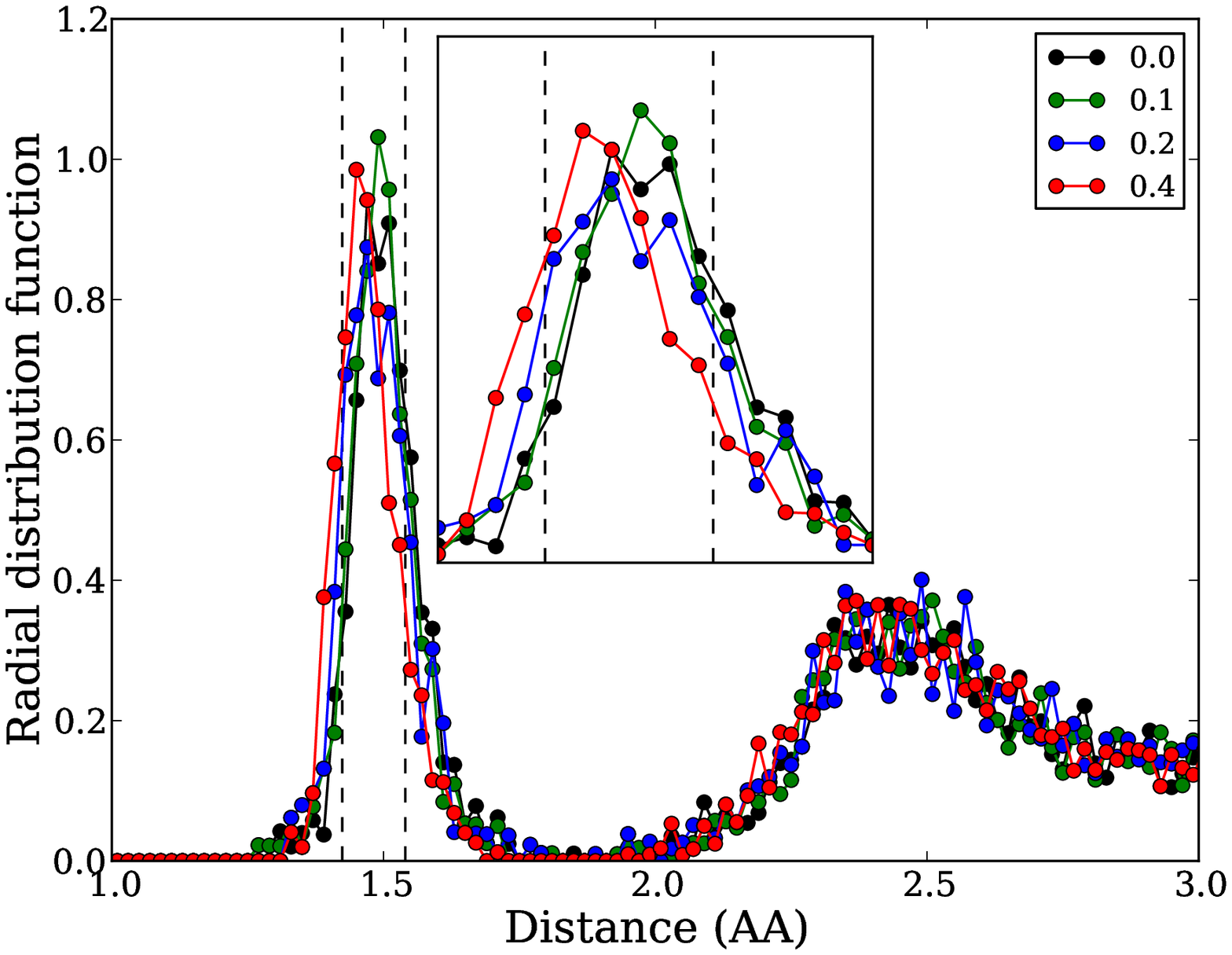}
    \caption{(Color online) Radial distribution functions of 3.4 g/cm$^3$ amorphous carbon with different constrained magnetization. Structures are relaxed under nonmagnetic (black lines) and constrained magnetization of 0.1 (green), 0.2 (blue), and 0.4~$\mu_B$/atom (red). Two vertical dashed lines indicate the optimized bond lengths of graphene (1.424~{\AA}) and diamond (1.540~{\AA}) using the same computational method. The inset shows the magnified view around the first peak.
    \label{fig:RDF}
}
\end{figure}

We compute the radial distribution function to quantitatively compare the structures of amorphous carbon with several different constrained magnetization (Fig.~\ref{fig:RDF}).
At zero magnetization (black curve), the peak position is close to the bond length of diamond (1.54~{\AA}), indicating the dominant $sp^3$ hybridization.
The density of 3.4~g/cm$^3$ here is slightly smaller than diamond (3.5~g/cm$^3$), but relatively dense compared with the normally observed amorphous carbon phase.
The  high density structure close to diamond, with 4-fold $sp^3$-like hybridization is naturally favored.

On the other hand, the peak position of radial distribution function moves toward the bond length of graphene (1.42~{\AA}) as we increase the constrained magnetization.
This corresponds to the fact that the number of 3-fold coordinated atoms must increase in order to have unpaired electrons which contribute to the magnetism.
The resulting 3-fold atom portions of zero, 0.1, 0.2, and 0.4~$\mu_B$/atom cases are 13, 19, 28, and 44~\%, respectively.
The bond length is close to 1.42~{\AA} when a large magnetization is imposed, again indicating that $sp^2$-like hybridization is necessary for the realization of spin polarization.
The change in the radial distribution is not significantly large, but the peak should be close to $sp^2$ bonding when amorphous carbon exhibits sizable amount of magnetic moment.

\subsection{Effect of density}
\begin{figure}[tbp]
  \includegraphics[width = 8.6cm]{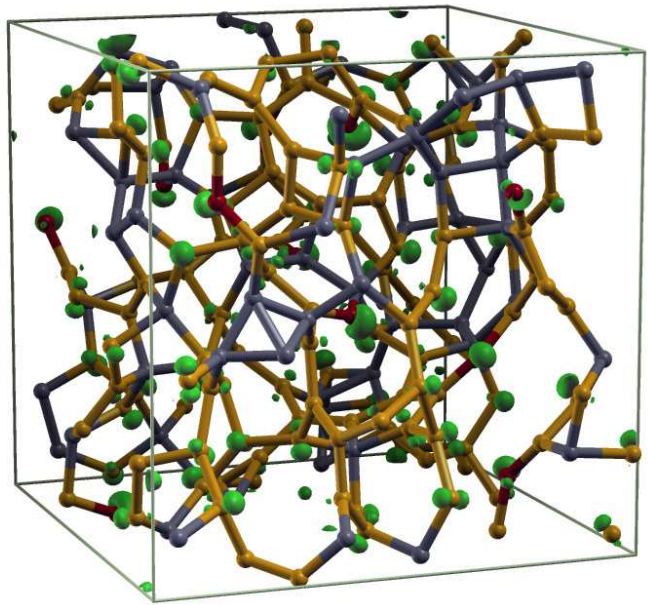}
    \caption{
    \label{fig:2604}
(Color online) Structure and spin charge density of amorphous carbon (density of 2.6~g/cm$^3$) with constrained magnetization of 0.4~$\mu_B$/atom. Red spheres represent 2-fold coordinated carbon atoms. 
}
\end{figure}

We also consider amorphous carbon with relatively low density.
Figure~\ref{fig:2604} shows the structure and spin charge density of the 2.6~g/cm$^3$ density and 0.4~$\mu_B$/atom case.
The structure shows more 3-fold coordinated atoms than in the high density 3.4~g/cm$^3$ case (see Fig.~\ref{fig:fig1}).
Another structural character is the appearance of 2-fold coordinated atoms (red spheres), which are not seen in the high density amorphous carbon.
The 2-fold and 3-fold portion here is 6~\% and 64~\% (22~\% higher than the previous case), respectively.
The appearance of 2-fold coordination and the increase in the 3-fold portion indicate the lower-coordination is not surprisingly favored in the low density case.
In fact, the 3-fold portion is 58~\% even when the system is not under a magnetic constraint.

The spin charge density is distributed on 2-fold and 3-fold coordinated carbon atoms, where 2-fold atoms have more unpaired electrons than 3-fold atoms.
The 3-fold atoms are a majority in this structure with most of the 3-fold atoms bonded to each other.
As discussed above, the remained $p$ orbitals in two bonded 3-fold atoms must be rotated relative to each other by 90$^\circ$ to avoid the formation of a $\pi$ bond.
This structural distortion increases the energy of amorphous carbon (358~meV/atom compared with the nonmagnetic case) although the formation of 3-fold coordinated atoms is favored in the low density case.

The total energies and 3-fold portion of amorphous carbon are summarized in Table~\ref{tb1}.
A high density amorphous carbon favors low 3-fold portion in both with and without magnetic constraints.
The energy difference between nonmagnetic and 0.4~$\mu_B$/atom constraint monotonically increases as a function of density (from 358 in 2.6~g/cm$^3$ to 578~meV/atom in 3.4~g/cm$^3$), indicating the difficulty of the formation of 3-fold coordinated atom.
Considering that the experimental $sp^3$ portion in Q-carbon is more than 75~\%,\cite{Narayan2015}, the density of Q-carbon should be around 3.2~g/cm$^3$ or denser. 
\begin{table*}
\caption{\label{tb1}
Total energy of amorphous carbon (meV/atom) with different densities and constrained magnetization. 
The energy is measured from the lowest total energy value of the nonmagnetic case with 3.0~g/cm$^3$ density. 
Values in parenthesis are the portion of 3-fold coordinated carbon atoms.
Each structure is obtained by independently relaxing the randomized atomic coordinates under each magnetic constraint.
}
\begin{ruledtabular}
    \begin{tabular}{ccccccc}
        Density (g/cm$^3$) &    2.6    &     2.8   &    3.0    &    3.2    &    3.4     \\
        \hline                                                    
        Nonmagnetic        & 48 (58~\%)& 70 (43~\%)&  0 (42~\%)&135 (35~\%)& 66 (13~\%) \\
        0.1 ($\mu_B$/atom) & 86 (57~\%)&133 (56~\%)& 53 (38~\%)&169 (31~\%)&101 (19~\%) \\   
        0.2 ($\mu_B$/atom) &155 (58~\%)&149 (51~\%)&233 (48~\%)&257 (30~\%)&318 (28~\%) \\
        0.4 ($\mu_B$/atom) &406 (64~\%)&420 (56~\%)&442 (51~\%)&550 (42~\%)&644 (44~\%)
    \end{tabular}
\end{ruledtabular}
\end{table*}

In general, a high magnetic constraint tends to require a high energy and large proportion of 3-fold sites.
The total energies of 0.4~$\mu_B$/atom structures are significantly higher than those without magnetic constraints. 
Interestingly, the relative energies of 0.1 and 0.2~$\mu_B$/atom is not substantially high when compared with the high energy for 0.4~$\mu_B$/atom.
For instance, the energy difference between nonmagnetic and 0.1~$\mu_B$/atom cases are 35~meV/atom even in amorphous carbon with a high density of 3.4 g/cm$^3$. 
Similarly, the energy is 122~meV/atom higher in the 0.2~$\mu_B$/atom and 3.2~g/cm$^3$ density case.
Considering the calculated energy difference between diamond and the lowest-energy amorphous carbon (3.0~g/cm$^3$) is 745~meV/atom, the energy difference here is relatively small.
Therefore, such a structure could be realized under the extreme synthesis condition used in creating amorphous Q-carbon.

\subsection{Releasing magnetic constraints}
\begin{figure}[tb]
  \includegraphics[width = 8.6cm]{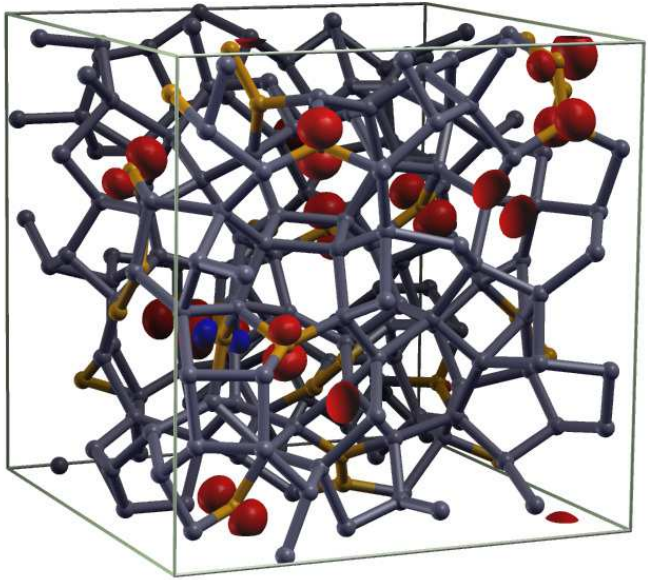}
    \caption{
    \label{fig:diff} 
(Color online) Spin charge density of amorphous carbon obtained with normal spin-polarized calculations without magnetic constraint (density of 3.4~g/cm$^3$). The initial structure is 0.1~$\mu_B$/atom constrained structure and relaxed without constraint. Red and blue isosurfaces represent majority and minority spin charge densities, respectively. See Fig.~\ref{fig:diff01} for comparison with the spin-constrained structure.
}
\end{figure}

We released the magnetic constraint to determine the nature of the magnetic moments without constraints.
The magnetization is reduced when we release the spin constraint and perform a standard spin-polarized calculation.
For example, the total magnetization and the sum of the absolute values of spin up and down moments are 0.044 and 0.051~$\mu_B$/atom, respectively in the 0.1~$\mu_B$/atom constraint and 3.4~g/cm$^3$ density case (see Fig.~\ref{fig:diff} for the spin charge densities).
The portion of 3-fold coordinated carbon atoms is also reduced from 19~\% to 14~\% as well because 4-fold coordination is more favored in high density amorphous carbon.
The total energy is 5~meV/atom higher when the same structure is calculated with spin unpolarized DFT, indicating a weak magnetic order among unpaired electrons.

The blue isosurface in Fig.~\ref{fig:diff} shows the minority spin charge density.
This implies the existence of finite antiferromagnetic order since the spin spontaneously becomes opposite in direction using a self-consistent calculation, even though we start the simulation with the same spin direction on each carbon atom. 
The distance between two carbon atoms with these two opposite spins is 2.24~{\AA} and they are separated by two 4-fold coordinated carbon atoms.
Here two $p$ orbitals have a slight overlap with each other, and this is believed to cause the antiferromagnetic order between the two spins.

\begin{figure}[tb]
  \includegraphics[width = 8.6cm]{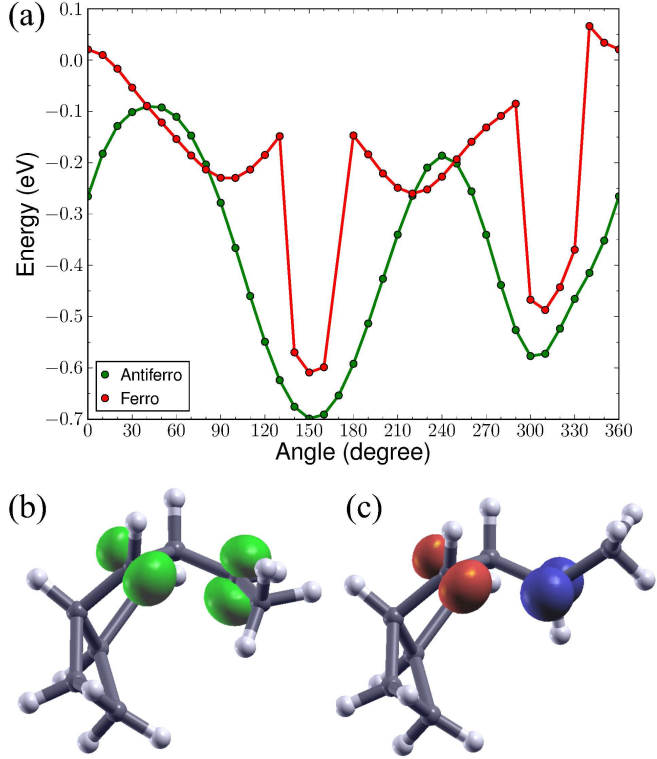}
    \caption{
    \label{fig:twist}
(Color online) (a) Relative energy as a function of twist angle of a $p$ orbital in a ``molecule" constructed from amorphous carbon. Green and red lines represent ferromagnetic and antiferromagnetic phases, respectively. The energy is measured from the total energy of the nonmagnetic phase at 0$^\circ$.
Structures and spin charge densities at twist angles of (b) 60$^\circ$ and (c) 150$^\circ$. Gray and white spheres represent carbon and hydrogen atoms. The structures are not relaxed and the only difference is the rotation of the $p$ orbital.
}
\end{figure}

To examine possible magnetic order between the opposite spins, we construct a ``molecule" by cutting out the amorphous structure around the two spins and terminate all dangling bonds with hydrogen atoms (C-H bond length is adjusted to 1~{\AA}).
This ``molecule" has 286~meV lower energy in the antiferromagnetic case than in the ferromagnetic case, similar to the fact that antiferromagnetic order spontaneously occurs in the amorphous structure.
However, the energy difference changes as we twist the angle of one $p$ orbital as described in Fig.~\ref{fig:twist}(a).
In general, the antiferromagnetic phase [the green line in Fig.\ref{fig:twist}(a)] has lower energy, but the ferromagnetic phase [the red line in Fig.~\ref{fig:twist}(a)] becomes more stable around 60$^\circ$ and 240$^\circ$.
For example, the energy is 43~meV lower than the antiferromagnetic case at 60$^\circ$.
The energy in the ferromagnetic phase is significantly low around 150$^\circ$ and 300$^\circ$ since the system prefers it to a nonmagnetic solution even when we start the simulation from a ferromagnetic initial condition.
Here the antiferromagnetic phase is in a relatively low energy state and at local minimum as opposite spin configuration is favored.

The structure and spin charge densities at the twist angles of 60$^\circ$ and 150$^\circ$ are illustrated in Figs.~\ref{fig:twist}(b) and \ref{fig:twist}(c).
The two $p$ orbitals are close to orthogonal and the overlap between the orbitals is limited in the 60$^\circ$ case [Fig.~\ref{fig:twist}(b)].
This orbital geometry enables these orbitals to be localized and ferromagnetic.
On the other hand, as we twist the orbital by 90$^\circ$, two $p$ orbitals are almost in the same plane and have substantial overlap with each other [Fig.~\ref{fig:twist}(c)].
Because of this large overlap, the system prefers opposite spin directions.
Although ferromagnetic phases are not in an energy local minimum with respect to the twist angle, we expect that such a geometry could appear and be a source of magnetism in amorphous carbon synthesized in an extreme condition.
%

\section{Summary}
In summary, unpaired electron in 3-fold $sp^2$-hybridized carbon atoms are necessary for producing magnetic behaviour in amorphous carbon.
We performed constrained magnetization MD and find that the portion of 3-fold atoms become large as we increase the magnetic constraints. 
In addition, those 3-fold atoms should (1) be isolated by 4-fold carbon atoms or (2) have 90$^\circ$ rotated $p$ orbitals when bonded to 3-fold atoms, to keep their unpaired electrons.
We also show that $p$ orbitals at 3-fold atoms in (1) could exhibit ferromagnetic (antiferromagnetic) order in amorphous carbon when they are close and orthogonal (in the same plane) to each other.
Our finding is useful for examining the magnetism and structure of Q-carbon, although the results presented in this paper is for general amorphous carbon systems.
We expect that the result presented in this work will be useful for explaining magnetic properties of amorphous carbon systems and be useful for designing new magnetic carbon materials.

\begin{acknowledgments}
YS and JRC acknowledge support from the U.S.~Department of Energy (DoE) for work 
on nanostructures from grant DE-FG02-06ER46286, and on algorithms by a subaward from the 
Center for Computational Study of Excited-State Phenomena in Energy Materials at 
the Lawrence Berkeley National Laboratory, which is funded by the U.S.~Department of Energy, 
Office of Science, Basic Energy Sciences, Materials Sciences and Engineering 
Division under Contract No.~DE-AC02-05CH11231, as part of the Computational Materials Sciences 
Program. Computational resources are provided in part by the National Energy Research 
Scientific Computing Center (NERSC) and the Texas Advanced Computing Center (TACC). 
MLC acknowledges support from the 
National Science Foundation Grant No.~DMR-1508412 and from the Theory of Materials Program at the 
Lawrence Berkeley National Lab funded by the Director, Office of Science and Office of Basic 
Energy Sciences, Materials Sciences and Engineering Division, U.S.~Department of 
Energy under Contract No. DE-AC02-05CH11231. MLC acknowledges useful discussions with Professor Jay Narayan.
\end{acknowledgments}

%
\end{document}